# Scaling-Up Model-Based-Development for Large Heterogeneous Systems with Compositional Modeling

**Christoph Herrmann**[1], **Holger Krahn**[2], **Bernhard Rumpe**[1], **Martin Schindler**[1], and **Steven Völkel**[1]

[1]Software Engineering Group, Department of Computer Science, RWTH Aachen, Germany

[2]Institute for Software Systems Engineering, TU Braunschweig, Germany

**Abstract**— *Model-based development and in particular MDA [1], [2] have promised to be especially suited for the development of complex, heterogeneous, and large software systems. However, so far MDA has failed to fulfill this promise to a larger extent because of tool support being inadequate and clumsy and methodologies not being appropriate for an effective development. This article discusses what went wrong in current MDA approaches and what needs to be done to make MDA suited for ultra-large, distributed systems.*

## 1. Introduction

MDA [1], [2] shall be especially suited for the development of complex, heterogeneous, and large software systems. However, it has so far failed to fulfill this promise to a larger extent because of missing appropriate tool support and methodologies. Especially the sequential development from abstract requirement models to an implementation with a number of (possibly) manual transformations hinders an effective development. The evolution of requirements that frequently occurs during long-lasting software development projects change artifacts on different levels of abstraction. Other artifacts must then co-evolve and stay consistent which is a major unsolved problem. In code-centric software development agile methods have proven to be more effective [3] than traditional waterfall-based approaches.

We therefore advocate an agile and compositional form of modeling which restricts the number of subsequent levels of abstractions by using an agile development process in which the artifacts are refined in a continuous fashion. The process requires a number of highly specialized and parameterized generators to fully automate the derivation of the final product.

The rest of the paper is structured as follows: Section 2 lists and discusses the problems when applying MDA with current technology. Section 3 explains how some of the problems can be solved when compositional modeling is used. Section 5 concludes the paper.

## 2. MDA problems

Figure 1 shows a typical form of an MDA approach. It breaks down the development process in a number of steps from informal requirements, through architecture, high- and low-level design, down to coding. Additionally present (yet not shown) are activities for planning the development in iterations and testing that can be model-driven. This standard MDA approach exhibits a number of problems:

1) The chain of models used from requirements down to the code contains too many sequential and manual steps. Normally, execution is only possible at the code level. This makes reviews and inspections of the higher-level models very important. However, these activities are manual and thus both: error prone and costly.
2) Evolution of a system primarily means evolution of the code. If models are not evolved for small changes, then larger adaptations cannot be model-based anymore, but are run on code basis only.
3) Tracing is proposed as a solution to keep models and code synchronized. However, tracing is expensive. Tracing the information between layers of models means considerably increased labor to establish the traced links. When evolving a system, evolving the model along the code does not only imply twice the effort to evolve both artifacts, but a significantly higher effort including the evolution of the trace and the review of its correctness.
4) A chain of models in an MDA project enforces either a sophisticated mega-tool that can handle everything or a smoothly integrated tool chain. However, both things currently do not work very well. A big monolithic tool on the one hand is not easy to produce and evolve. Therefore, these tools are either very hard to adapt to changing technology or domain evolution or do not really provide domain or technology specific capabilities. On the other hand tool chains need a smooth integration, possibly with a tracing of dependencies back and forth. If these tools are not prepared for integration right away, it is terribly hard to actually couple them, even if the coupling is loose, e.g., through transformation of model formats only. Such a loose coupling is state of the art today, enabling replacement, e.g., the early end for requirements modeling and the late end for code generation in a relatively handsome way [4]. Still, a replacement of a tool or even only a tool update that affects the model format or meaning of modeling elements lead to larger effort necessary to handle the tool chain.

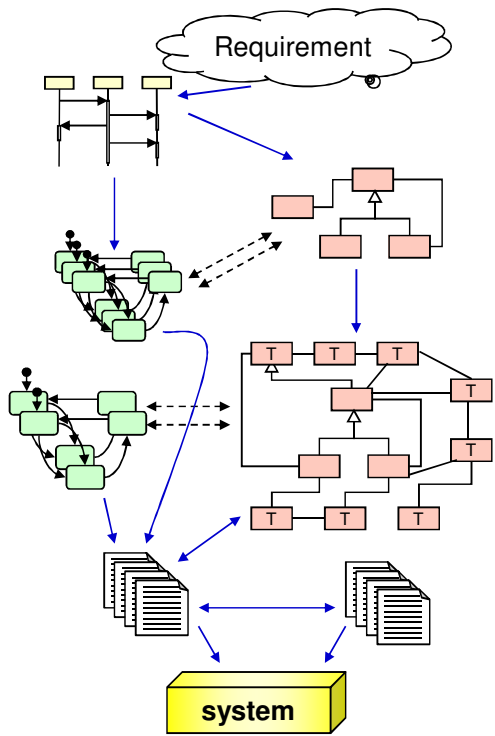

| | |
|---|---|
| | use cases and scenarios: sequence diagram: users viewpoint |
| | application classes (PIM) |
| | state machines |
| | class diagram Nr. 2 („PSM"): adaptation, extension, technical design + behavior for technical classes |
| | code generation + integration with manually written code |
| | complete and running system |

Fig. 1: Exemplary development process using MDA

5) Last but not least monolithic code generation is a big obstacle for an efficient development. Currently code generation is not incremental, but whenever a developer changes a single element (say an attribute) the whole set of classes in the model is updated. This largely prevents agile and efficient development. Instead, people tend to generate code very rarely thus thinking and developing at the model level without the very helpful feedback from execution.

MDA in its current form has not yet delivered what it has promised. Applying MDA or similar approaches to ultra-large systems will exhibit further deficiencies. We expect the following issues to raise up:

6) MDA projects are rather plan driven. They try to use top down development of a system as a total. Ultra-large distributed systems will high-likely not be developed by one unit or company, but will be developed and installed by a consortium. Each site might run their own project to add nodes to the overall system. E.g., clinical information systems or cross-company production systems are built this way. As the projects are on an individual basis, a top-down master plan does not exist. MDA cannot be applied as is, but external interfaces, exchange formats, etc. have to be defined, both on implementation as well as on modeling level. When interfaces become standard, these interfaces even become part of the requirements analysis.

7) Developers over company or even state boundaries will not trust each other in developed code and components very much and thus a common approach of development needs to be split into a separated, cooperative approach where interfaces are shared, while internal parts of software are added in local subprojects.

8) Quality of Service, reliability, and other important artifacts of the overall distributed system will become difficult to plan and maintain, if the nodes are developed independently.

9) Evolution of such systems over boundaries will be especially hard and can probably only be reached in a smooth and down-ward compatible way.

While some of the issues discussed above are immanent to MDA and need to be handled by appropriate setup of a development process, others can be handled by developing standards especially for interfaces. Such a standard does not only deal with the interfaces between independently developed components, but also with common evolution of those interfaces for new and enhanced data. Those interfaces are far beyond technical encodings, like XML or SOA [5], but domain specific, e.g., for health care or banking. However, those interface standards define syntactic structures as well as transmission formats and connection protocols. For those structures and protocols we need appropriate modeling languages, either on basis of the UML or domain specific languages to describe the common interfaces and cut real code in heterogeneous technical spaces through varying generators. Corba's IDL is a first attempt into this approach [6].

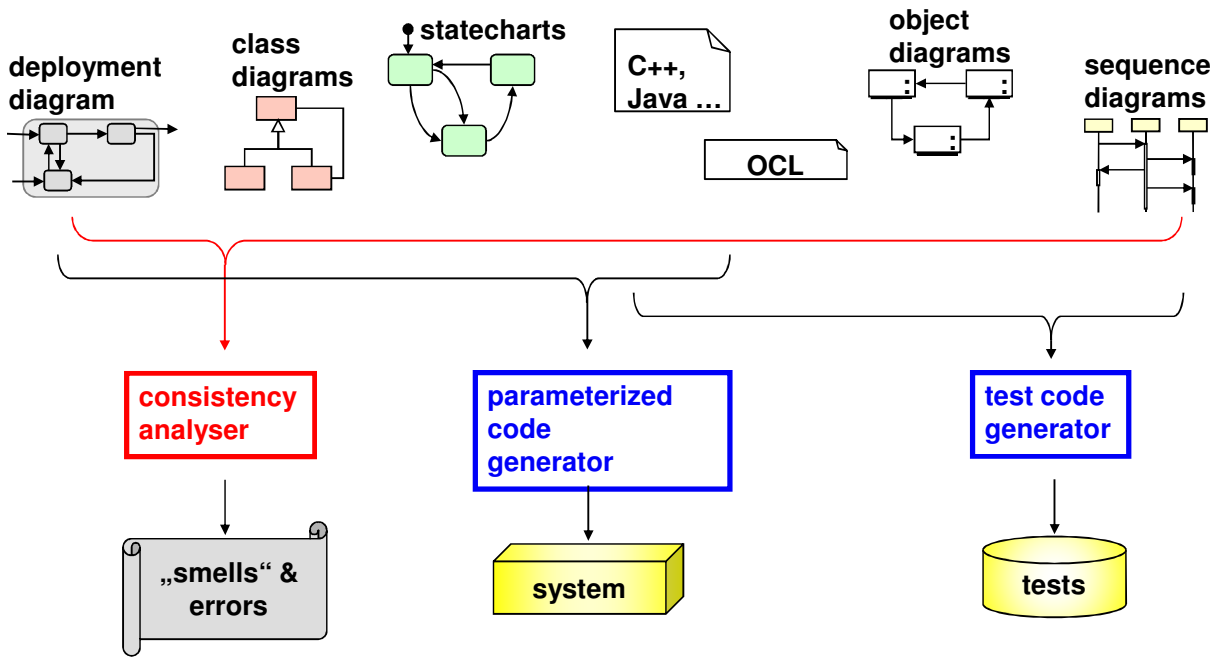

Fig. 2: Modular compositional modeling process

We also feel that one important issue of the development of large heterogeneous systems is the possibility to develop agile, yet model driven. As argued, MDA is anything else but agile [7]. In the following section we therefore concentrate solely on this issue.

## 3. Compositional Modeling

Much has been written about composition in the context of components (e.g. [8], [9]) and of specific specification languages, like algebraic specification (e.g. [10], [11]). For programming languages modularity is common since Parnas article [12]. In programming languages modularity has a number of nice consequences. A modular component written in a programming language is a rather self-contained unit that allows

- independent understanding, of what the software unit does,
- rather independent reasoning on the properties of the interfaces,
- independent compilation of the source, and
- late loading and binding of the independently compiled units.

While the former points are important for understanding, maintaining, and evolving large systems, the latter points are important for an agile development process. Independent compilation means that not everything has to be recompiled when one local change was made. It also means that pre-compiled components and frameworks can be developed, shipped to users, and integrated into systems. Composition of components therefore does not only have impact of independent understanding and development, but also on independent compilation and deployment. This technique of late binding is common to individual classes as well as groups of coherent classes (often called components) and was possible through the definition of clear syntactic interfaces.

In [13] we have discussed different forms of composability of models. One of them argues the transfer of this approach to MDA. Here we need "syntactic interfaces" between models, the possibility for independent compilation, and late binding of generated code. Instead today, if in one class one attribute is changed, tools usually generate everything again. Even worse, instead of concentrating on subsets of currently modified diagrams, tools tend to maintain the overall "complete model" including all diagrams and using the complete set of diagrams for regeneration. This consumes too much time, in particular as the code generators are not efficiency optimized and usually generate source code that is then recompiled. Thus developers tend to rarely push the re-generation button.

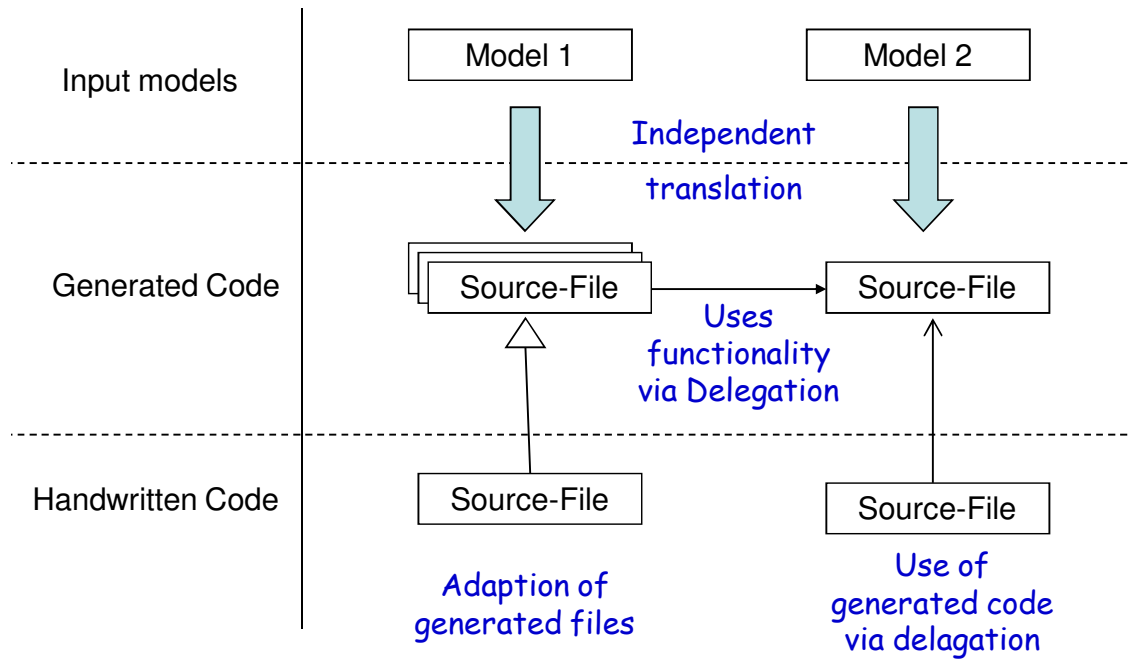

Fig. 3: Modular code generation

We therefore propose an improved approach of the use of models, where all models are incrementally transformed to code and the generated pieces of code are bound together very late in the process. This process is depicted graphically in Figure 2. It also includes the agile development of automated tests using additional models that are specifically suited for exemplary descriptions. Of course not everything is being worth being modeled. Therefore, we assume normal code in Java or C++ being integrated in addition.

We propose that these models used for code generation are rather redundancy free. In particular, we do not use round trip engineering, where code and models are just two different viewpoints.

## 4. Compositional Code Generation

The code generator is an especially important concept in this approach: It needs to be capable of generating self-contained pieces of code, usually complete classes that can be compiled independently (see Figure 3). Thus, if a single model is changed, only the code for that model needs to be regenerated.

This e.g. means that class diagrams do not generate code frames that are then manually filled. In our approach a class diagram is instead used to generate the data structure including association management and access functions, the data base connection, web presentation, and other interfaces to neighbor systems. All generated classes are complete and cannot be modified, but its attributes can be accessed appropriately and subclasses can be built. Each state machine is generated into one (sometimes more) separate classes that are integrated through appropriate delegation mechanisms. We are currently enhancing this approach and have experienced that the use of appropriate design pattern helps quite a lot to come up with compositional models.

Experiences show that although the type system of the target language does make it more difficult, it is possible to generate compositionally. However, we were sometimes forced to modify the provided Java source code according to some rules as well. In particular it was necessary to replace static constructs, such as static method call and attributes and new-constructs. The new construct, e.g., is consistently replaced by factory calls. Such a factory is generated from the appropriate class diagram automatically and since the Java/C++-sources are modified automatically as well, they are transparent and the user does not have to deal with.

## 5. Conclusion

In summary, we experienced that modular compositional modeling is possible and can be efficiently and transparently be used to generate code. If the system is heterogeneous and complex and thus needs independent development, such a compositional approach is a necessary prerequisite to decompose the task and develop in parallel. If the overall system is ultra distributed and therefore necessarily developed by independent parties, models can serve as standardized interfaces where the additional, local models and code are written against. Developers of this additional code do not need to know internal technical details for protocols etc., but those are added (composed) implicitly by the code generator, which is to be provided in addition to the standardized interface models.

**Acknowledgement:** The work presented in this paper is undertaken as a part of the MODELPLEX project. MODELPLEX is a project co-funded by the European Commission under the "Information Society Technologies" Sixth Framework Programme (2002-2006).